\documentclass[%
 reprint,
superscriptaddress,
 amsmath,amssymb,
 aps,
reqno]{revtex4-1}

\usepackage{graphicx}
\usepackage{dcolumn}
\usepackage{bm}
\usepackage{float}
\usepackage[caption=false]{subfig}
\usepackage{amsmath}
\usepackage{natbib}
\begin{document}


\title{Direct Laser Ion Acceleration and Above-Threshold Ionization at Intensities from $10^{21}$ W/cm$^{2}$ to $3 \times 10^{23}$ W/cm$^{2}$}


\author{A. Yandow}
\affiliation{%
Center for High Energy Density Science \\
2515 Speedway RLM 12.204\\
Austin, TX 78712}
 
\author{T. Toncian}
\affiliation{%
Center for High Energy Density Science \\
2515 Speedway RLM 12.204\\
 Austin, TX 78712}%

\affiliation{%
Institute of Radiation Physics \\
 Helmholtz-Zentrum Dresden-Rossendorf \\
  01328 Dresden, Germany
}%

\author{T. Ditmire}%
\affiliation{%
Center for High Energy Density Science \\
2515 Speedway RLM 12.204\\
 Austin, TX 78712}%

\date{\today}

\begin{abstract}
Calculations on the dynamics of ions and electrons in near-infrared laser fields at intensities up to $3 \times 10^{23}$ W/cm$^2$ are presented. We explore the acceleration of ions in a laser focus by conservation of canonical momentum during ionization events and by the ponderomotive force in the \textit{f}/1 focal geometry required to reach such intensity. At intensity exceeding 10$^{23}$ W/cm$^2$, highly charged ions are expelled from the laser focus before they can interact with the laser pulse at peak intensity, decreasing the predicted ionization yields of deeply-bound states. We consider the interaction of a tightly-focused, \textit{f}/1 laser pulse with krypton at an intensity of $3 \times 10^{23}$ W/cm$^{2}$ and a pulse duration of 140 fs. We find that the ions and electrons are accelerated to energies in excess of 2 MeV/nucleon and 1.4 GeV, respectively. Ponderomotive expulsion of the parent ions decreases the total number of ultra-relativistic ATI electrons produced by tunneling ionization from the K-shell states of krypton but does not change their energy spectrum.
\end{abstract}

\pacs{Valid PACS appear here}
\maketitle


\section{\label{sec:level1} Introduction}

Above-threshold ionization (ATI), first observed by Agostini et al. in 1979 \cite{Agostini1979}, is the fundamental response of an atomic system to a strong laser field and the dominant laser energy absorption mechanism in low-density plasmas. When the density of photons becomes high enough, ATI can be treated as a quasi-classical, two-step process: the bound electron is freed by adiabatic tunneling and the continuum dynamics can be found by integrating the Lorentz force equations. The two-step model of ATI has been used to explain high-harmonic generation in gases \cite{Corkum1994}\cite{Krause1992} and non-sequential double ionization (NSDI) \cite{Fittinghoff1992}\cite{Watson1997}. Measurements of the ATI electron energy spectrum and angular distribution provide direct evidence that tunneling ionization dominates in infrared laser fields above $10^{15}$ W/cm$^{2}$ \cite{Corkum1989a}\cite{McNaught1998}. Laser intensities currently exceed $2 \times 10^{22}$ W/cm$^{2}$ \cite{Tiwari2019}\cite{Yoon2019} and now approach  $10^{23}$ W/cm$^{2}$, a regime where ATI offers a method of accelerating electrons to GeV energies over a few microns. 
	
Free electron dynamics in near-infrared laser fields become relativistic at intensity above $10^{18}$ W/cm$^{2}$. The trajectories of electrons liberated by tunneling ionization in these fields are folded forward by the laser magnetic field, and the electrons are observed to gain momentum in the laser forward direction \cite{Moore1995a}. Despite the onset of relativistic free electron motion, precision measurements of the ionization rates of argon at intensity up to $2 \times 10^{19}$ W/cm$^{2}$ have agreed with the nonrelativistic Ammosov-Krainov-Delone/Perelomov-Popov-Terent’ev (ADK/PPT) tunneling model of ionization \cite{Chowdhury2001}. Although ionization channels involving inelastic re-scattering are suppressed by relativistic electron drift in the continuum \cite{Dammasch2001}, it is less clear how relativistic laser-electron interactions affect the primary ionization process. The laser magnetic field is thought to play a role in stabilizing bound states by giving the electron a nonzero momentum component transverse to the tunneling path, effectively increasing the height of the tunneling barrier \cite{Yakaboylu2013}. However, numerical treatment of ionization as an ensemble of classical electron orbits demonstrated the inclusion of the laser magnetic field has negligible effects on the ionization rates at intensity up to $10^{23}$ W/cm$^{2}$ \cite{Grugan2012}. Tunneling rate models for hydrogen-like ions that use Dirac wavefunctions predict the ionization rate will be $\sim$1/3 of the nonrelativistic rate above $10^{23}$ W/cm$^{2}$ \cite{Milosevic2002}, but the corrections from the laser magnetic field are negligible \cite{Milosevic2002a}. Precision measurements of highly charged ion yields and ATI electron energy spectra are needed to verify the magnitude of expected relativistic corrections. 

In this article we numerically investigate the ionization and electron dynamics in ultra-intense laser fields that will be available with hybrid OPCPA/Nd:glass systems scheduled to come online in the near future \cite{Rus2017}. We employed numerical methods that allowed us to model tunneling ionization and calculate ATI electron energy spectra without neglecting the motion of the parent ions in the laser field. The ions will acquire energy from their interaction with the laser field, which has been numerically explored as a method for creating high-energy proton pulses for cancer therapy \cite{Salamin2008}. When laser intensity exceeds $10^{23}$ W/cm$^{2}$ in the \textit{f}/1 focal geometry required to reach this intensity with a 10-PW class laser, the ponderomotive force expels the ions before they can interact with strongest  laser field. For highly-charged krypton ions in a laser focus, we demonstrate that ponderomotive ion expulsion will substantially reduce the number of K-shell ionization events. Simulations of the ATI electron dynamics show evidence of two dominant electron acceleration mechanisms for the highest-energy ATI electrons. We conclude that direct laser ion acceleration (DLIA) will necessitate the development of novel experimental techniques to measure ionization rates when intensity exceeds $10^{21}$ W/cm$^{2}$.

\section{\label{sec:level1} Ion Dynamics}
	The physics of direct laser ion acceleration (DLIA) is analogous to the acceleration of ATI electrons in non-relativistic, near infrared laser fields. We can extend the quasi-classical two-step model of Corkum et al. \citep{Corkum1989a} to approximate ion energy contributions from two sources: residual drift that arises from conservation of canonical momentum in the laser field at each ionization event and from ponderomotive acceleration. Assuming a sequential tunneling ionization process, nonrelativistic ion motion, and negligible ponderomotive force on the ion, we can express the energy as  
\begin{equation}
E_{\text{drift}} \approx \frac{1}{2 m \omega^{2}}\Bigg[ \sum_{q=1}^{q_{\text{max}}} E(t_{q}) sin(\phi_{q})\Bigg]^{2} 
\end{equation}in a linearly polarized laser field, where $E(t_{q})$ is the laser electric field amplitude and $\phi_{q}$ is the laser phase at the time of the $q^{\text{th}}$ ionization event. The ion residual drift energy spectrum can only be calculated numerically as the energy spectrum includes information about the laser phase and field strength at every ionization event, and therefore the energy spectrum depends strongly on the target ion species and its electronic shell structure. 

	For the tight \textit{f}/1 focal geometries considered in this paper the ponderomotive energy contribution dominates the residual drift energy at intensities above $10^{21}$ W/cm$^{2}$. One can approximate the ponderomotively ejected ion energy as 

\begin{equation}
\begin{aligned}
E_{\text{pond}} \approx \frac{1}{2m}\Bigg[\int\limits_{-\infty}^{\infty} \nabla U_{p}(\textbf{x}(t), t, q(t)) dt\Bigg
]^{2} \\ = \frac{1}{32m^{3}\omega^{4}}\Bigg[\int\limits_{-\infty}^{\infty} q(t)^{2}\nabla \textbf{E}(\textbf{x}(t), t)^2 dt\Bigg
]^{2}
\end{aligned}
\end{equation}
where $U_{p} = q^{2}E^{2}/4m\omega^{2}$ is the ion ponderomotive potential and q(t) is the ion charge as a function of time. The details of ponderomotive ion ejection are complex, and the final energy depends on charge state history, laser intensity, spot size, pulse duration, and initial position within the laser focus, and thus no general formula for final ion energy exists. Simulations of the ion dynamics illustrate both residual drift and ponderomotive DLIA when considering laser-ion interactions on 10-PW class laser systems.

	We simulate the ion dynamics by numerically integrating the Lorentz force equations using the Runge-Kutta-Felberg method. We neglect space charge fields in our simulations and they can be ignored when considering skimmed effusive atomic beams with densities on the order of $10^{11}$ cm$^{-3}$ and a diameter of $\sim$ 1 mm. The atom is neutral before arrival of the laser pulse and, at each timestep, the probability of ionization is calculated using the ADK/PPT tunneling ionization rate \cite{Ammosov1986}\cite{Perelomov1966}. We assume the single active electron approximation, neglecting collective tunneling and re-collision effects, and increment the charge state using straightforward Monte-Carlo methods. The ion charge state histories can be reproduced using ionization potentials from Kelly and Harrison \cite{Kelly1971} for krypton and the NIST Atomic Spectral Database for all other elements \cite{NIST_ASD}. We model the laser field as a Gaussian focus and we include non-paraxial corrections up to fifth order in the diffraction angle \cite{Salamin2007}. 
	
	Figures 1 and 2 present simulated ion energy spectra for argon and krypton. The peak intensities are  chosen to be the barrier suppression intensity (BSI) \cite{Augst2008} of the hydrogen-like ion charge states. Numerical solution of the time-dependent Schr\"{o}dinger equation demonstrates that the ADK/PPT model likely overestimates the tunneling probability as the barrier suppression regime is approached \cite{Bauer1999}. However, for highly-charged ions, the probability of ionization by tunneling will be significant before the barrier suppression intensity is reached, so these corrections can typically be ignored \cite{Ciappina2019}. The estimated pulse duration below which barrier suppression effects must be considered is on the order of 5 fs for the ionization of Ar$^{17+}$ \cite{Kostyukov2018}, much shorter than the 140 fs pulse duration we consider throughout this article. The ADK/PPT model and the single electron approximation may be less accurate when considering the ionization events in the low-intensity leading edge of the laser pulse, but they will provide a better estimate of the charge state history than assuming a pre-ionized target. We randomly choose the initial position of the neutral atom within the laser confocal region and propagate the laser pulse through, repeating for $10^{5}$ trials. The laser central wavelength is 1057 nm throughout this article. 
	
\begin{figure}[!]

\begin{minipage}{\linewidth}

\subfloat[]{\label{main:a}\includegraphics[width = 0.9\linewidth]{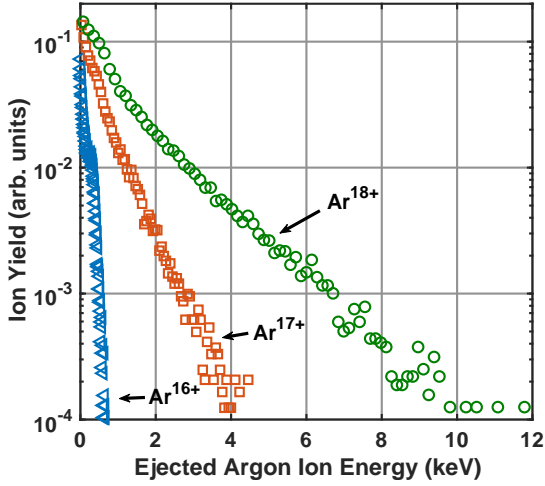}} 

\subfloat[]{\label{main:b}\includegraphics[width = 0.9\linewidth]{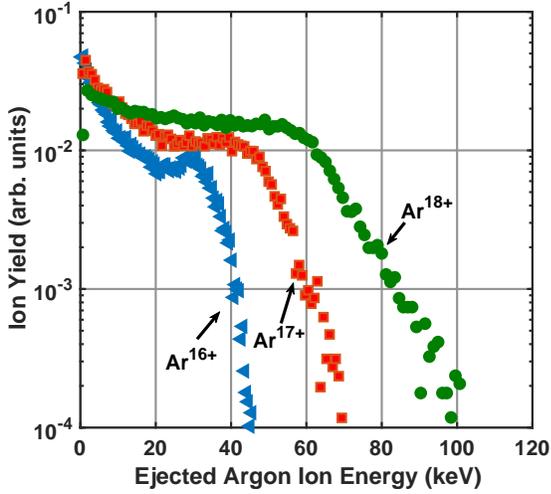}}

\subfloat[]{\label{main:c}\includegraphics[width = 0.75\linewidth]{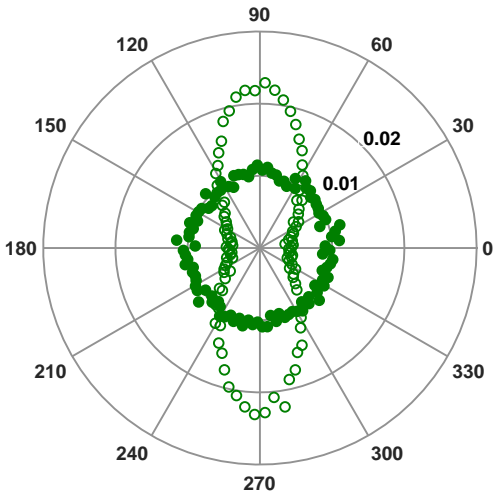}}

\end{minipage}

\caption{Energy spectra of argon ions ejected from a laser focus with peak intensity of $4.7 \times 10^{21}$ W/cm$^{2}$ from a) a 30 $\mu$m focal diameter (open markers) and b) 3 $\mu$m focal diameter (closed markers). Green circles, red squares, and blue triangles represent Ar$^{16+}$, Ar$^{17+}$, and Ar$^{18+}$, respectively. c) Comparison of angular distribution (arb.) of Ar$^{18+}$ in both geometries}.	
		
\end{figure}		
		
\begin{figure}[t!]
\includegraphics[width = \linewidth]{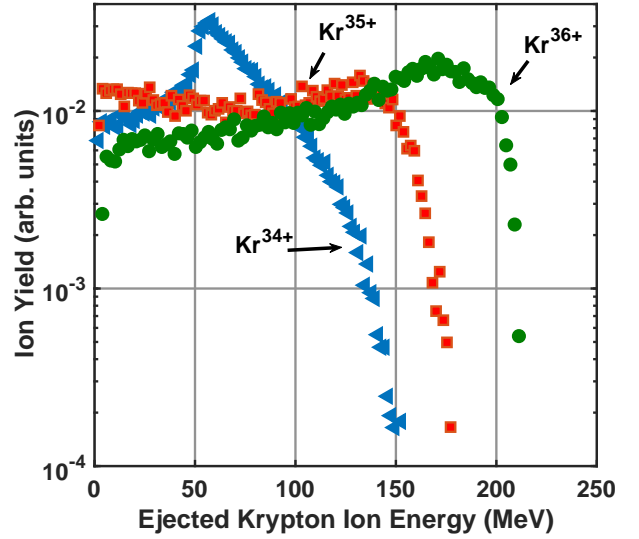}
\caption{Energy spectra of krypton ions ejected from the laser focus with peak intensity of $3 \times 10^{23}$ W/cm$^{2}$ and a 3 $\mu$m focal diameter. Green circles, red squares, and blue triangles represent Kr$^{34+}$, Kr$^{35+}$, and Kr$^{36+}$, respectively.  The peak around 60 MeV for Kr$^{34+}$ is an artifact of the simulation boundaries, which exclude the large volume of lower energy Kr$^{34+}$ produced several Rayleigh ranges away from the focal plane.}

\end{figure}

	Figure 1a and 1b show the calculated energy spectra of argon in \textit{f}/10 and \textit{f}/1 focal geometries, which we take to have 1/$e^{2}$ diameters of 30 $\mu$m and 3 $\mu$m, respectively. The peak laser intensity is $4.7 \times 10^{21}$ W/cm$^{2}$, which can be attained by a 10 PW-class laser system in the larger focus. Residual drift ion energy dominates in the \textit{f}/10 geometry for Ar$^{17+}$ and Ar$^{18+}$, yielding an exponential distribution of ejected ion energies. The large gap in calculated ion energy between Ar$^{16+}$ and the higher charge states reflects the large difference in ionization potential (and hence, the BSI) between the L-shell and K-shell electrons of argon. The ponderomotive force dominates the dynamics of ions expelled from the \textit{f}/1 focus, and ion energy reflects the strength of the ponderomotive force at each ion’s initial position in the focus. Residual drift is necessary to explain the hot ion tail of Ar$^{18+}$ in the \textit{f}/1 geometry. Figure 1c shows the azimuthal angular distribution of Ar$^{18+}$ ejected from each focus, confirming that ions accelerated by residual drift will be expelled preferentially along the laser polarization direction while ponderomotively accelerated ions will be expelled radially. Figure 2 shows the energy spectrum of krypton ions expelled from an \textit{f}/1 focus where the peak laser intensity is $3 \times 10^{23}$ W/cm$^{2}$. The peak krypton ion energy is in excess of 200 MeV. We will show that, in this ultra-intense regime, the highest-energy ejected ions will gain nearly the full ponderomotive energy.	

\begin{figure}[t!]
\includegraphics[width = \linewidth]{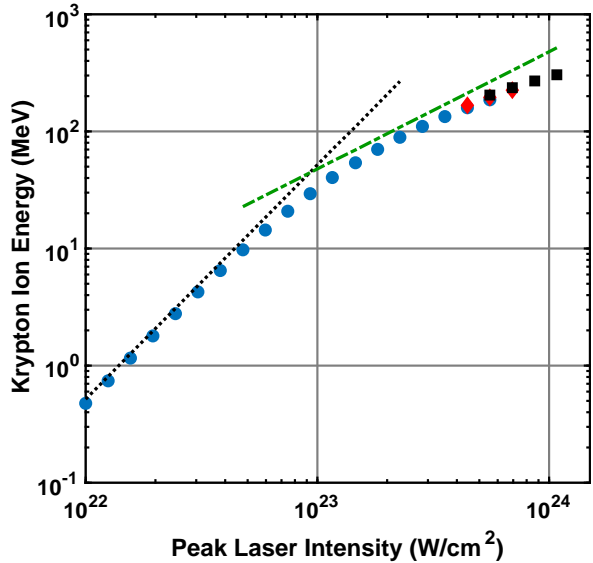}
\caption{Average energy of 500 krypton ions originating from ($w_{o}$/2, 0, 0) for different peak laser intensities. Peak laser intensity is defined at (0, 0, 0). 
Blue circles indicate Kr$^{34+}$, red diamonds indicate Kr$^{35+}$, and black squares indicate Kr$^{36+}$. The black dotted line and the green dashed line are the short pulse maximum energies (Eq. 3) or ponderomotive energy for a Kr$^{34+}$ ion, respectively. The focal spot 1/e$^{2}$ diameter is 3 $\mu$m.} 
\end{figure}	

	Ponderomotive DLIA can be divided into long and short pulse regimes. In both regimes, the ion experiences a ponderomotive force $f_{p} = - \nabla U_{p} \sim -U_{p}/w_{o}$. When the laser pulse duration is much shorter than the timescale on which the ion leaves the focus, a small amount of the ponderomotive potential is converted to kinetic drift energy. Assuming a stationary ion, integrating the ponderomotive force over the laser pulse duration $\tau_{p}$ gives an impulse $\Delta p \sim -\frac{U_{p} \tau_{p}}{w_{o}}$, yielding a quadratic ion energy scaling with peak intensity. A calculation assuming a Gaussian spatial mode and temporal profile as well as a constant ion charge q gives a maximum energy of
	
\begin{equation}
E_{ion, sp} = \frac{\pi}{8\text{ln(2)Exp(1)}} \frac{q^{4}}{\omega^{4} c^{2} \epsilon_{o}^{2} m^{3}} (\frac{I_{o} \tau_{\text{p}}}{w_{o}} )^{2}
\end{equation}

due to ponderomotive acceleration in the short pulse regime, where $\tau_{\text{p}}$ is the intensity full-width at half maximum (FWHM) pulse duration, $I_{o}$ is the peak laser intensity, and $w_{o}$ is the beam waist. When the laser pulse duration becomes comparable to the ejection timescale (long pulse regime), we expect a significant fraction of the ion ponderomotive energy, which scales linearly with intensity, is converted to kinetic drift energy. An approximate ion ejection timescale $\tau_{ej} \approx w_{o}\sqrt{2m/U_{p}}$ can be derived from the ponderomotive model. For an f/1 focus, this timescale is on the order of 1 ps for hydrogen-like argon at its BSI but is $\sim 120$ fs for hydrogen-like krypton at its BSI.

	The mechanisms of DLIA discussed in this article closely resemble the nonrelativistic picture of ATI, with the key difference that the ponderomotive dynamics often dominate residual drift even in the short-pulse regime. Residual drift plays a minor role because the ion charge-to-mass ratio does not change much during ionization events occurring at the peak laser field strength and the canonical momenta gained during successive ionization events can cancel each other. The interplay between ponderomotive acceleration and residual drift acceleration depends strongly on the ion species and its trajectory, so the transition between residual drift and short-pulse ponderomotive dynamics can only be predicted numerically. The ponderomotive impulse model presented for the short-pulse ponderomotive regime closely parallels the “surfing” picture presented by Bucksbaum et al., where the ponderomotive potential was observed to accelerate (or decelerate) nonrelativistic ATI electrons crossing the focal volume of a laser field between the ATI electron source and detector \cite{Bucksbaum2008}.

	We confirmed the ponderomotive ejection timescale estimate by simulating five hundred krypton ions originating from the point of strongest ponderomotive force in a laser focus with a 1/e$^{2}$ diameter of 3 $\mu$m. The averages of these ion energies are presented in Figure 3. As the peak laser intensity in the focus increases, the ponderomotive force on the ion also increases. When intensity exceeds $10^{23}$ W/cm$^{2}$, the ions are expelled from the focus on the timescale of the laser pulse duration, gaining a significant fraction of the ion ponderomotive energy at its initial position. This short to long pulse regime transition causes the change in ion energy scaling with intensity from quadratic to linear seen in Fig. 3 at $10^{23}$ W/cm$^{2}$. In the short pulse ponderomotive DLIA regime, the ions originating from the point of strongest ponderomotive force in the focus will have the highest energy, and we observe agreement with Eq. 3. As $10^{24}$ W/cm$^{2}$ is approached the scaling becomes sub-linear because the ion is expelled from the focus before the peak intensity is reached.
	
\begin{figure}[b!]
\includegraphics[width = \linewidth]{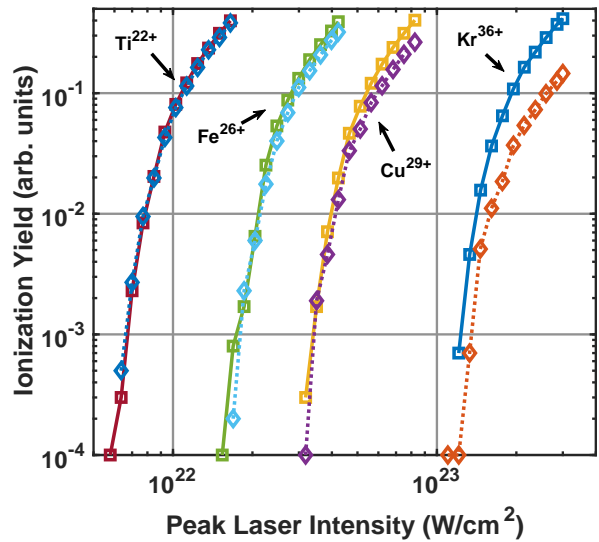}
\caption{Ionization probability for hydrogen-like states of ions (left-to-right: Ti$^{21+}$, Fe$^{25+}$, Cu$^{28+}$, and   Kr$^{35+}$) in a laser focus as laser intensity increases above $10^{23}$ W/cm$^{2}$, including (diamonds, dotted curves) and excluding (squares, solid curves) ion motion. Color included for clarity. The focal spot 1/e$^{2}$ diameter is 3 $\mu$m.}
\end{figure}		
	
	Calculating ionization yields in the long-pulse DLIA regime is complicated by the fact that the number of ions in the focal volume does not remain constant over the laser pulse duration. The ions will be expelled from the laser focus before being ionized further by the peak strength of the laser field, and we expect calculations including and excluding ion motion to differ substantially when $\tau_{p}$ $\sim$ $\tau_{ej}$. We calculate the ionization yields by integrating $10^{4}$ initial atom positions distributed over a fixed focal volume, bounded by the iso-intensity shell where the probability of K-shell ionization is greater than 0.05 for stationary ions. Simulated ion yields including and excluding the ion motion are given in Fig. 4. An increasing difference between the stationary and mobile ion models is observed as laser intensity approaches $10^{23}$ W/cm$^{2}$. Mobile and immobile yields for hydrogen-like krypton differ by a factor of $\sim$3, demonstrating ion motion must be accounted for when calculating ionization yields. We expect that the ion yield decrease due to ponderomotive ion ejection will be comparable to the decrease caused by relativistic corrections to the ionization rates.

\section{\label{sec:level1} Electron Dynamics}
	The final energies of ATI electrons depend strongly on the laser phase at the instant of ionization \cite{McNaught1998}. In the ultra-relativistic regime (I $>$ $10^{21}$ W/cm$^{2}$), the energy spectrum of ATI electrons is also sensitive to target ion position in the laser focus. Numerical studies of ATI electrons produced by the ionization of hydrogen-like argon suggest the highest-energy electrons originate from ions located from the front and sides of the confocal region \cite{Gordon2017}\cite{Pi2015}. If the most energetic ATI electrons originate from the edges of the focal volume, where the ponderomotive force on the ions is strongest, we expect a disproportionate reduction in the number of high-energy ATI electrons produced because their parent ions will be ejected from the focus before ionization. We therefore simulated the dynamics of ATI electrons produced by the ionization of hydrogen-like highly-charged ions in the laser focus.

	The electron initial conditions were generated by simulating the ion dynamics in the focal volume using the same method for calculating the ionization yields described in the previous section. Each ion velocity, position, and time is recorded at the instant the hydrogen-like state is ionized. The electrons are born into the laser field at rest with respect to their parent ions. Unless explicitly mentioned otherwise the effects of ion motion are included when calculating the initial conditions. The Coulomb field is neglected after ionization and the electron dynamics are found by integrating the equations of motion using a modified 7th-order Dormand-Prince scheme \cite{Prince1981}. The equations of motion, given below, are the Lorentz force equation with a correction added for the leading term of the Landau-Lifshitz radiation friction force \cite{Tamburini2010}\cite{Tamburini2012}

\begin{equation}
\begin{aligned}
\frac{d{\textbf{p}}}{dt} = -e\bigg(\textbf{E} + \frac{\textbf{v}}{c} \times \textbf{B}\bigg) \\ + \frac{2}{3} r_{c}^{2} \Bigg\{ \bigg[  \bigg(\textbf{E} + \frac{\textbf{v}}{c} \times \textbf{B}\bigg) \times \textbf{B} + \bigg(\frac{\textbf{v}}{c}  \cdot \textbf{E}\bigg) \textbf{E} \bigg] \\ - \gamma^{2}\bigg[\ \bigg(\textbf{E} + \frac{\textbf{v}}{c} \times \textbf{B}\bigg)^{2}-\bigg(\frac{\textbf{v}}{c}  \cdot \textbf{E}\bigg)^{2}\bigg]\frac{\textbf{v}}{c} \Bigg\}
\end{aligned}
\end{equation}

where $r_{c}=e^{2}/m_{e}c^{2}$ is the classical electron radius and $\gamma$ is the Lorentz factor. The radiation reaction correction decreases the final energy of the highest-energy electron we simulated by approximately $0.2\%$.  $10^{4}$ unique electron initial conditions were used for each simulation. The laser polarization is oriented along the x-axis and the laser forward direction is oriented along the postive z-axis. 

	Figure 5 shows nearly identical ATI electron energy spectra for from ionization of hydrogen-like krypton at an intensity of 3 x 10$^{23}$ W/cm$^{2}$ when ion motion is ignored (closed markers ) or included (open markers) when generating the initial electron conditions. ATI simulations were performed over two decades of laser intensity and the parent ion positions at the moment of ionization were recorded. Figure 6 shows the ion positions from which the $10\%$ most energetic ATI electrons originate, projected onto the XZ plane, for three ion species (Ar$^{17+}$, Ti$^{21+}$, and Kr$^{35+}$). The peak laser intensities in the Gaussian focus and temporal envelope are chosen to be their respective barrier suppression intensities. The color (shape) of the markers indicates the direction of the paraxial laser field $E_{x}$ at the moment of ionization. A transition in the electron acceleration mechanisms is evidenced by the development of three distinct, compact regions in the focus where the high-energy electrons are generated. At an intensity of $3 \times 10^{23}$ W/cm$^{2}$, the vast majority of the high-energy ATI electrons originate from ions concentrated along the laser axis in the back of the confocal region where the ponderomotive force is weaker than it is at the beam waist. These ions are not likely to be expelled more quickly than other ions in the focus, explaining the consistency between the simulations excluding and including ion motion seen in Fig. 5.

\begin{figure}[t!]
\includegraphics[width = \linewidth]{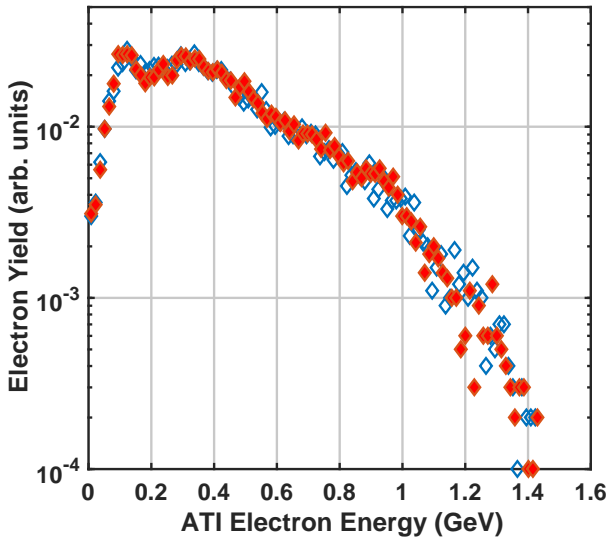}
\caption{ATI electron energy spectra for electrons produced by ionization of Kr$^{35+}$ integrated over the focal volume. Electron initial conditions are calculated by simulation of immobile (closed diamonds, red) or mobile (open diamonds, blue) krypton ions in the focus. The peak laser intensity is $3 \times 10^{23}$ W/cm$^{2}$ and focal spot 1/e$^{2}$ diameter is 3 $\mu$m.} 
\end{figure}			


\begin{figure}[!]

\begin{minipage}{0.78\linewidth}

\subfloat[]{\label{main:a}\includegraphics[width = \linewidth]{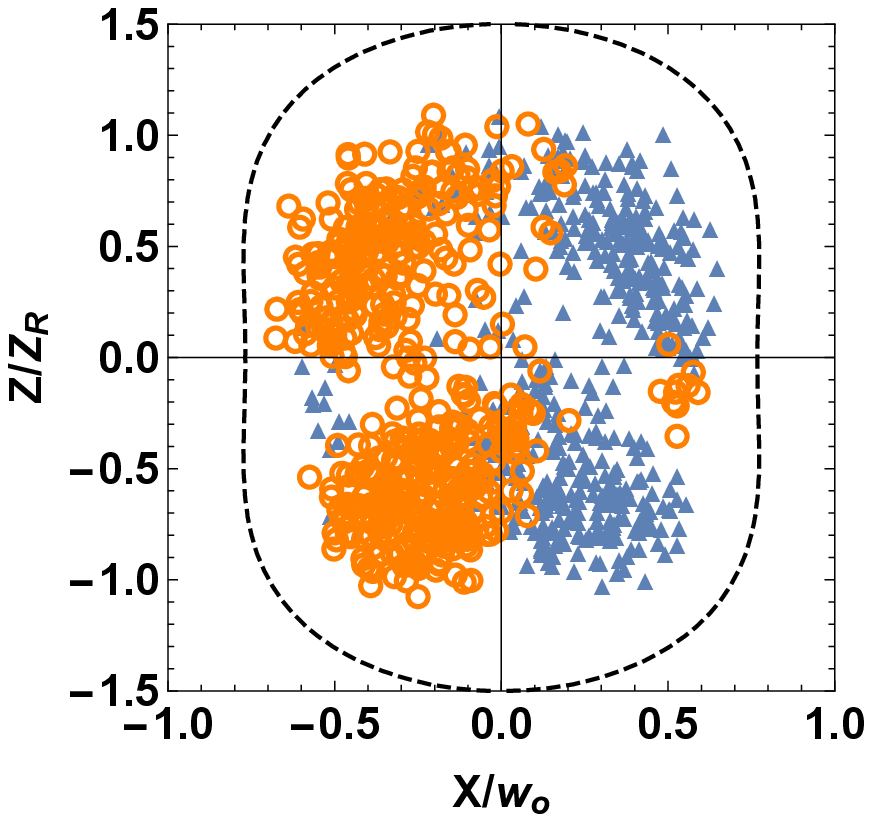}} 

\subfloat[]{\label{main:b}\includegraphics[width = \linewidth]{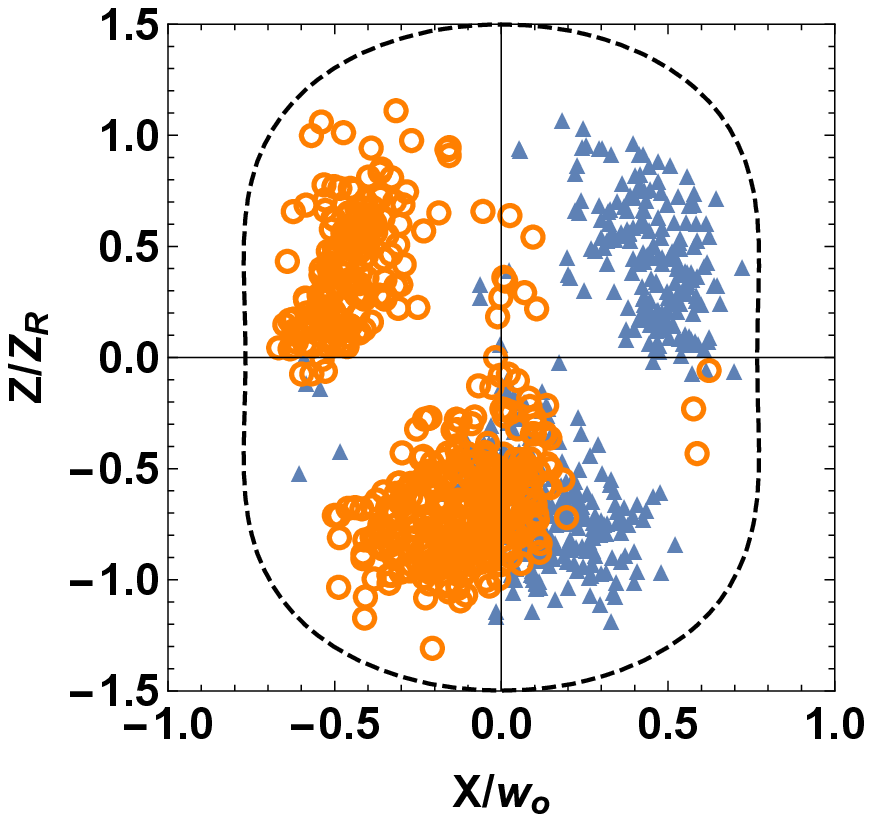}}

\subfloat[]{\label{main:c}\includegraphics[width = \linewidth]{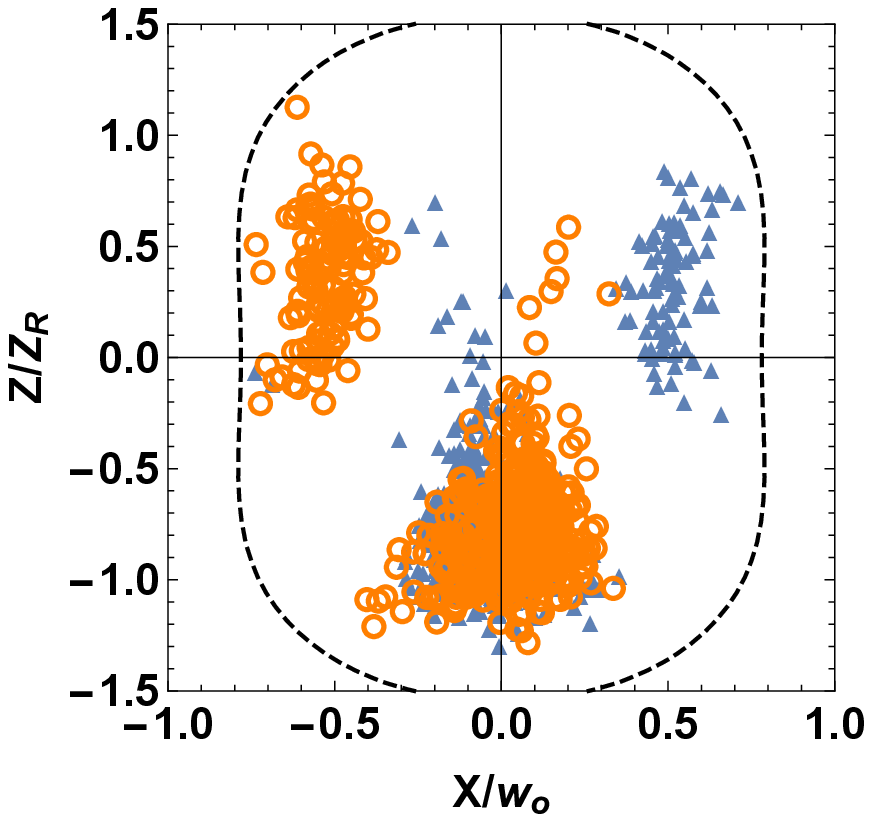}}

\end{minipage}
\caption{Initial positions of the $10\%$ highest-energy ATI electrons produced by a) Ar$^{17+}$ at 4.7 $\times 10^{21}$ W/cm$^{2}$, b) Ti$^{21+}$ at 1.6 $\times 10^{22}$ W/cm$^{2}$, and c) Kr$^{35+}$ at 3 $\times 10^{23}$ W/cm$^{2}$. Open orange circles (color online) denote negative $E_{x}$, solid blue triangles denote positive $E_{x}$. The focal spot 1/e$^{2}$ diameter is 3 $\mu$m. The black dashed curve represents the focal volume boundary containing all atoms (before arrival of the laser puslse) in the simulation.}
\end{figure}

	The presence of these three distinct regions leads us to identify two ultra-relativistic electron acceleration mechanisms where the longitudinal electric field $E_{z}$ plays an important role. The first-order non-paraxial laser fields are oriented along the laser propagation direction and must be included to ensure the electric and magnetic laser fields remain divergence-free and therefore satisfy Maxwell’s equations. The first-order nonparaxial magnetic field was shown to be essential for a correct description of the ponderomotive force by Quesnel and Mora, but the first-order nonparaxial electric field plays little role in the electron dynamics at softly relativistic intensity \cite{Quesnel1998}. However, $E_{z}$ will do work on the electron in the ultra-relativistic regime, when it points nearly parallel to the electron velocity. The work provided by the longitudinal electric field substantially reduces the calculated energies of ultra-relativistic ATI electrons because it dephases these electrons from the paraxial laser electric field \cite{Maltsev2003a}. We refer to the mechanism accelerating electrons from the back of the confocal region as rephasing acceleration (RA) and the mechanism accelerating electrons out of the front of the focus as direct injection acceleration (DIA). 

Electrons accelerated by either mechanism accelerate to nearly the speed of light in the laser forward direction in a small fraction of a laser cycle. Although the magnitude of the nonparaxial electric field directed along the laser propagaton direction is lower than the paraxial fields by an order of magnitude, it contributes substantially to the rate of electron energy change $dE/dt$ $\alpha$ $\textbf{E} \cdot \textbf{v}$ as $v_{z}/v_{x} \approx \sqrt{\gamma/2}$ \cite{Maltsev2003a}. The trajectories of two test electrons, one accelerated by DIA and one by RA, were calculated as they exited the confocal region nearly parallel to the z-axis. Figure 7a shows the Lorentz factor $\gamma$ for the RA electron (solid blue) and the DIA electron (dashed-dotted green) with all fields included. The same trajectories with all nonparaxial fields excluded with the exception of the first-order nonparaxial magnetic field are displayed for comparison (dashed red for the RA mechanism, dotted black for the DIA mechanism). Dephasing from the paraxial field in this scenario is driven only the superluminal phase velocity of the focused beam \cite{Popov2008}, leading to lower electron energy gain and loss rates. Figure 7b shows both the normalized $E_{x}$ (no markers) and $E_{z}$ (open squares) experienced by the two test electrons when the nonparaxial corrections are included. The signs of the fields in Fig. 7b provide a qualitative description of the acceleration mechanisms.

\begin{figure}[t!]

\begin{minipage}{0.9\linewidth}

\subfloat[]{\label{main:a}\includegraphics[width = \linewidth]{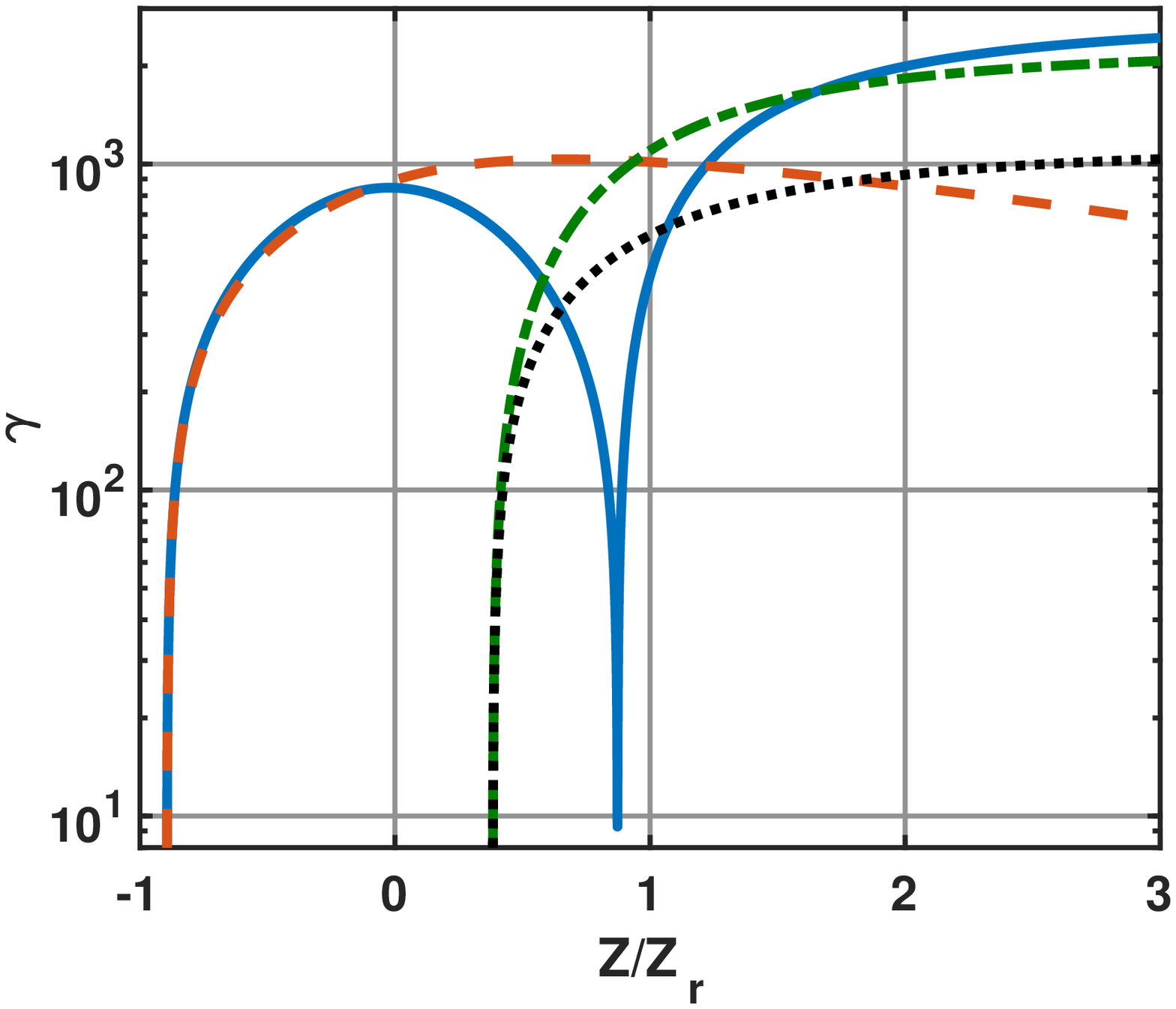}} 

\subfloat[]{\label{main:b}\includegraphics[width = \linewidth]{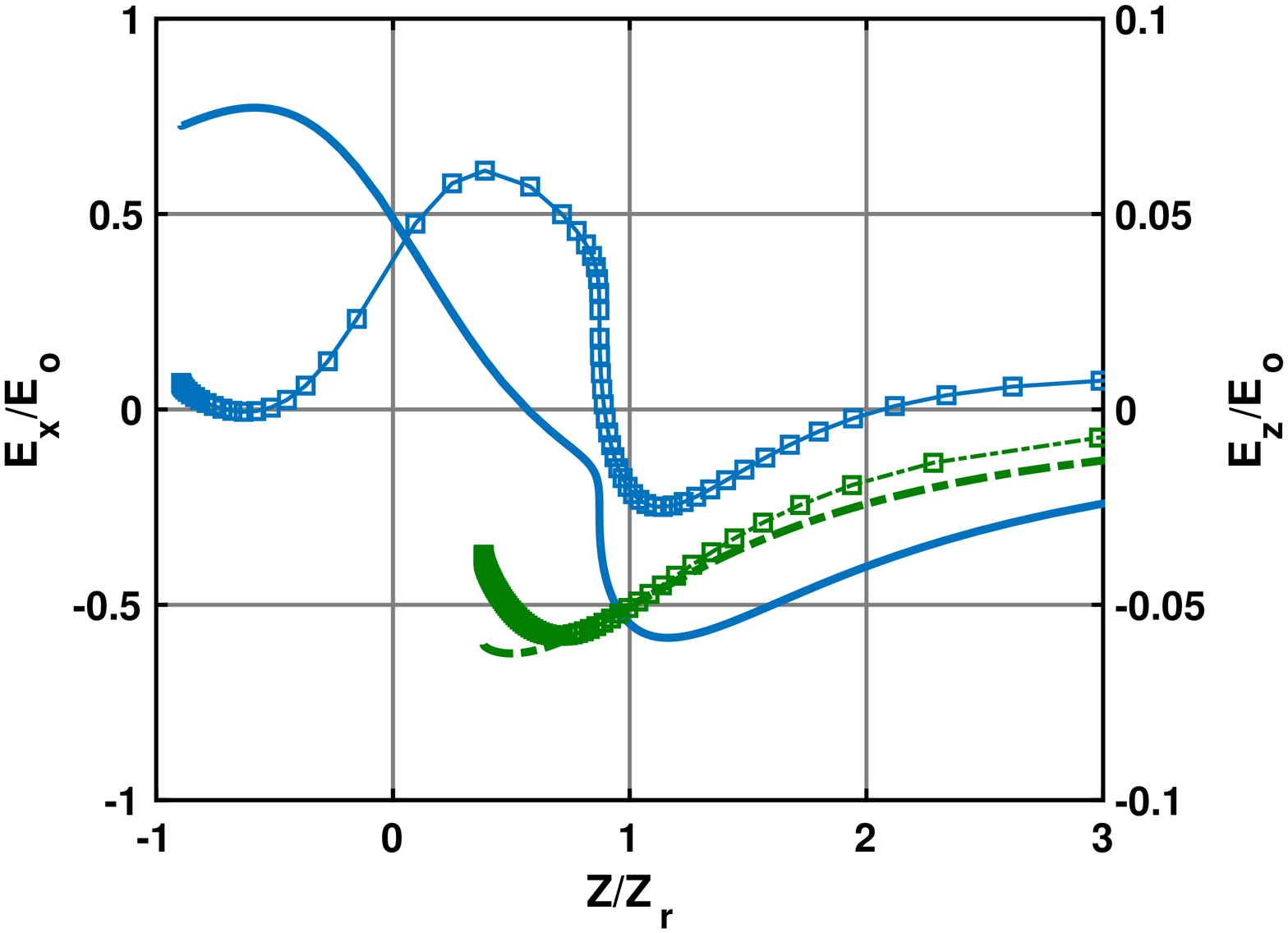}}

\end{minipage}
\caption{a) Test electron $\gamma$ as it travels through focus for representative initial conditions for RA (solid blue, negative z origin) and DIA (dashed-dotted green, positive z origin). Red dashed lines and black dotted lines represent identical initial conditions with the only first-order nonparaxial magnetic correction included. b) Electric fields acting on the test electrons. Color scheme is the same as a) for $E_{x}$ (left scale). $E_{z}$ (right scale) is represented by corresponding line style with open square markers. Electron motion is very nearly parallel to the z-axis.}
\end{figure}

	For the DIA mechanism, the electron is born in the front of the confocal region. The sign of $E_{x}$, as noted in Fig 6c, immediately accelerates the electron toward the z-axis in both DIA regions while $E_{z}$ is negative, as seen in Fig 7b. The paraxial magnetic field folds the electron trajectories toward the laser forward direction, and the electron remains nearly in phase with the paraxial laser field as it gains energy from $E_{z}$. In the RA mechanism, the electron loses energy to work done by $E_{z}$ faster than it gains energy from work done by $E_{x}$, and begins to decelerate before the sign of $E_{x}$ changes. Complete deceleration is averted by the longitudinal magnetic field, which exerts a force ($\alpha$ $v_{\perp}B_{z}$) in the XY plane, reversing the electron motion in the x-direction so it can gain energy from $E_{x}$ without completely decelerating. These electrons gain more energy than the initially stationary DIA electrons "born" in the front of the confocal region because they gain energy at a higher rate from the paraxial field ($dE/dt$ $\alpha$ $c E_{x}$) due to their relativistic velocity at the instant of acceleration. 

\section{\label{sec:level1} Discussion}

	While our observation of the highest-energy ATI electrons originating from the back of the confocal region is unexpected, it does not necessarily contradict the results of previous studies. We do not find qualitative agreement with Pi et al. for origin positions of the highest energy ATI electrons from Ar$^{17+}$ \cite{Pi2015}, but we emphasize that final electron energy is much more sensitive to the laser phase at the moment of ionization than it is to initial position in the focus. The over-the-barrier ionization model used in their paper will yield a different distribution of initial phases than the quantum ADK model, which can make direct comparison of our results difficult. We also restricted our analysis of initial electron positions to an intensity regime the ion yield is not strongly saturated. At higher intensities where the ion yield is strongly saturated, electrons will originate from a larger volume of the focus and the distribution of preferred initial positions may change. Other numerical studies of free electron acceleration by few-cycle, petawatt, radially polarized pulses \cite{Marceau2012} and fully nonparaxial linearly polarized pulses \cite{Popov2008} show high-energy scattered electrons originating from positions the back of the confocal region.

	DLIA poses a formidable challenge to ionization rate experiments at intensity above $10^{21}$ W/cm$^{2}$. Even in focal geometries where ponderomotive acceleration can be neglected, canonical momentum conservation accelerates the ions to keV energies. Wiley-McLaren time-of-flight methods, which were used to measure ionization yield in virtually all previous ionization rate experiments with noble gases \cite{Chowdhury2001}\cite{Augst1989}\cite{Augst2008}, will not be able to capture all the ions nor will they have sufficient resolution to resolve the closely-spaced charge-to-mass ratio peaks when the ions have the broad energy spectra we predict. Ionization yield measurements for hydrogen-like and helium-like charge states of argon and krypton will have to be inferred from the high-energy ATI electrons ejected from the laser focus. ATI electrons generated by tunneling from the K-shell will typically have at least an order of magnitude higher peak energy than ATI electrons originating from the L-shell. Ponderomotive expulsion of ions from the laser focus will not disproportionally reduce the number of high-energy ATI electrons produced when the ion yield is not saturated. 

	The techniques recently proposed by Ciappina et al. \cite{Ciappina2019} to determine peak laser intensity by measuring relative yields of highly-charged ions produced in the focus will lose substantial intensity resolution if the only experimental observable is the relative ATI electron yields from different atomic shells rather than relative yields of ion charge states within those shells. Further systematic study of ATI electron energies produced by each charge state will be necessary to determine the limit of intensity resolution and will aid the selection of appropriate target atoms for different intensities. The rate equation model of ionization will overestimate the yield of high charge states, and thus underestimate the laser intensity, when the long-pulse DLIA regime is reached.
	
	DLIA will substantially reduce the number of K-shell ionization events in a tightly-focused laser beam at intensities greater than 10$^{23}$ W/cm$^{2}$, and must be included to accurately calculate the expected number of ionization events. Shorter laser pulses ($<$ 25 fs) can mitigate the effect of DLIA on the ionization yield, but ionization yields do not compare well with indirect intensity measurements inferred from the focal spot size and pulse duration diagnostics due to B-integral accumulation in the diagnostic transport \cite{Akahane2006}. For longer pulses of hybrid OPCPA/Nd:glass laser systems, the intensities inferred from ionization yields and indirect measurements compare favorably \cite{Link2006}, and measuring ionization threshold intensities above 10$^{23}$ W/cm$^{2}$ with a longer pulse system can enable accurate cross-calibration of shorter pulse systems. The electronic shell structure of the target atoms can also be studied by examining modulations in the angular distribution and energy spectrum of the ATI electrons \cite{Dichiara2008}\cite{Ekanayake2013}, but these features will be washed out by the rapid intensity ramp-up of shorter pulses.
\section{\label{sec:level1}Conclusion}

	A complete understanding of ion and ATI electron dynamics in tightly-focused, nonparaxial laser fields will be essential to study relativistic corrections to the tunneling ionization rates at intensities above $10^{23}$ W/cm$^{2}$. At these intensities, broad DLIA energy spectra ($E_{max}$ $\sim$ ~ 2 MeV/nucleon) will make direct detections of ion charge states and measurement of the ionization yields extremely challenging. Future experimental studies of ionization rates at such intensity will require the development of large-area, high dynamic-range electron detectors capable of detecting individual 100 MeV-1.5 GeV electrons expelled from the laser focus. 
	
	This work was supported by the Air Force Office of Scientific Research (AFOSR) awards FA9550-14-1-0045 and FA9550-17-1-0264, and by the Jane and Mike Downer Endowed Presidential Fellowship in Laser Physics in Memory of Glenn Bryant Focht.





\bibliography{PRA_Simulation_Paper.bib}

\begin{thebibliography}{42}
\providecommand{\natexlab}[1]{#1}
\providecommand{\url}[1]{\texttt{#1}}
\expandafter\ifx\csname urlstyle\endcsname\relax
  \providecommand{\doi}[1]{doi: #1}\else
  \providecommand{\doi}{doi: \begingroup \urlstyle{rm}\Url}\fi

\bibitem[Agostini et~al.(1979)Agostini, Fabre, Mainfray, Petite, and
  Rahman]{Agostini1979}
P.~Agostini, F.~Fabre, G.~Mainfray, G.~Petite, and N.~K Rahman.
\newblock {Free-Free Transitions Following Six-Photon Ionization of Xenon
  Atoms}.
\newblock \emph{Phys. Rev. Lett.}, 42\penalty0 (17):\penalty0 1127--1130, 1979.

\bibitem[Corkum(1993)]{Corkum1994}
Paul~B Corkum.
\newblock {Plasma Perspective on Strong-Field Multiphoton Ionization}.
\newblock \emph{Phys. Rev. Lett.}, 71\penalty0 (13):\penalty0 1994--1997, 1993.
\newblock ISSN 0022-0957.
\newblock \doi{10.1093/jxb/46.11.1693}.

\bibitem[Krause et~al.(1992)Krause, Schafer, and Kulander]{Krause1992}
Jeffrey~L. Krause, Kenneth~J. Schafer, and Kenneth~C. Kulander.
\newblock {High-order harmonic generation from atoms and ions in the high
  intensity regime}.
\newblock \emph{Phys. Rev. Lett.}, 68\penalty0 (24):\penalty0 3535--3538, 1992.
\newblock ISSN 00319007.
\newblock \doi{10.1103/PhysRevLett.68.3535}.

\bibitem[Fittinghoff et~al.(1992)Fittinghoff, Bolton, Chang, and
  Kulander]{Fittinghoff1992}
D.~N. Fittinghoff, P.~R. Bolton, B.~Chang, and K.~C. Kulander.
\newblock {Observation of nonsequential double ionization of helium with
  optical tunneling}.
\newblock \emph{Phys. Rev. Lett.}, 69\penalty0 (18):\penalty0 2642--2645, 1992.
\newblock ISSN 00319007.
\newblock \doi{10.1103/PhysRevLett.69.2642}.

\bibitem[Watson et~al.(1997)Watson, Sanpera, Lappas, Knight, and
  Burnett]{Watson1997}
J~B Watson, A~Sanpera, D~G Lappas, P~L Knight, and K~Burnett.
\newblock {Nonsequential double ionization of helium}.
\newblock \emph{Phys. Rev. Lett.}, 78\penalty0 (10):\penalty0 1884--1887, 1997.
\newblock ISSN 10797114.
\newblock \doi{10.1103/PhysRevLett.78.1884}.

\bibitem[Corkum et~al.(1989)Corkum, Burnett, and Brunel]{Corkum1989a}
P.~B. Corkum, N.~H. Burnett, and F.~Brunel.
\newblock {Above-threshold ionization in the long-wavelength limit}.
\newblock \emph{Phys. Rev. Lett.}, 62\penalty0 (11):\penalty0 1259--1262, 1989.
\newblock ISSN 00319007.
\newblock \doi{10.1103/PhysRevLett.62.1259}.

\bibitem[McNaught et~al.(1998)McNaught, Knauer, and Meyerhofer]{McNaught1998}
S.~J. McNaught, J.~P. Knauer, and D.~D. Meyerhofer.
\newblock {Photoelectron initial conditions for tunneling ionization in a
  linearly polarized laser}.
\newblock \emph{Phys. Rev. A - At. Mol. Opt. Phys.}, 58\penalty0 (2):\penalty0
  1399--1411, 1998.
\newblock ISSN 10941622.
\newblock \doi{10.1103/PhysRevA.58.1399}.

\bibitem[Tiwari et~al.(2019)Tiwari, Gaul, Martinez, Dyer, Gordon, Spinks,
  Toncian, Bowers, Jiao, Kupfer, Lisi, McCary, Roycroft, Yandow, Glenn,
  Donovan, Ditmire, and Hegelich]{Tiwari2019}
G.~Tiwari, E.~Gaul, M.~Martinez, G.~Dyer, J.~Gordon, M.~Spinks, T.~Toncian,
  B.~Bowers, X.~Jiao, R.~Kupfer, L.~Lisi, E.~McCary, R.~Roycroft, A.~Yandow,
  G.~D. Glenn, M.~Donovan, T.~Ditmire, and B.~M. Hegelich.
\newblock { Beam distortion effects upon focusing an ultrashort petawatt laser
  pulse to greater than $10^{22}$ W/cm$^{2}$ }.
\newblock \emph{Opt. Lett.}, 44\penalty0 (11):\penalty0 2764, 2019.
\newblock ISSN 0146-9592.
\newblock \doi{10.1364/ol.44.002764}.

\bibitem[Yoon et~al.(2019)Yoon, Jeon, Shin, Lee, Lee, Choi, Kim, Sung, and
  Nam]{Yoon2019}
Jin~Woo Yoon, Cheonha Jeon, Junghoon Shin, Seong~Ku Lee, Hwang~Woon Lee, Il~Woo
  Choi, Hyung~Taek Kim, Jae~Hee Sung, and Chang~Hee Nam.
\newblock {Achieving the laser intensity of $5.5 \times 10^{22}$ W/cm$^{2}$
  with a wavefront-corrected multi-PW laser}.
\newblock \emph{Opt. Express}, 27\penalty0 (15):\penalty0 20412, 2019.
\newblock \doi{10.1364/oe.27.020412}.

\bibitem[Moore et~al.(1995)Moore, Knauer, and Meyerhofer]{Moore1995a}
C~I Moore, J~P Knauer, and D~D Meyerhofer.
\newblock {Observation of the Transition from Thomson to Compton Scattering in
  Multiphoton Interactions with Low-Energy Electrons}.
\newblock \emph{Phys. Rev. Lett.}, 74\penalty0 (13):\penalty0 2439--2442, 1995.

\bibitem[Chowdhury et~al.(2001)Chowdhury, Barty, and Walker]{Chowdhury2001}
Enam~A. Chowdhury, C.~P.J. Barty, and Barry~C. Walker.
\newblock {"Nonrelativistic" ionization of the L-shell states in argon by a
  "relativistic" $10^{19}$ W/cm$^{2}$ laser field}.
\newblock \emph{Phys. Rev. A - At. Mol. Opt. Phys.}, 63\penalty0 (4):\penalty0
  1--6, 2001.
\newblock ISSN 10502947.
\newblock \doi{10.1103/PhysRevA.63.042712}.

\bibitem[Dammasch et~al.(2001)Dammasch, D{\"{o}}rr, Eichmann, Lenz, and
  Sandner]{Dammasch2001}
Matthias Dammasch, Martin D{\"{o}}rr, Ulli Eichmann, Ernst Lenz, and Wolfgang
  Sandner.
\newblock {Relativistic laser-field-drift suppression of nonsequential multiple
  ionization}.
\newblock \emph{Phys. Rev. A - At. Mol. Opt. Phys.}, 64\penalty0 (6):\penalty0
  4, 2001.
\newblock ISSN 10941622.
\newblock \doi{10.1103/PhysRevA.64.061402}.

\bibitem[Yakaboylu et~al.(2013)Yakaboylu, Klaiber, Bauke, Hatsagortsyan, and
  Keitel]{Yakaboylu2013}
Enderalp Yakaboylu, Michael Klaiber, Heiko Bauke, Karen~Z. Hatsagortsyan, and
  Christoph~H. Keitel.
\newblock {Relativistic features and time delay of laser-induced tunnel
  ionization}.
\newblock \emph{Phys. Rev. A - At. Mol. Opt. Phys.}, 88\penalty0 (6):\penalty0
  1--21, 2013.
\newblock ISSN 10502947.
\newblock \doi{10.1103/PhysRevA.88.063421}.

\bibitem[Grugan et~al.(2012)Grugan, Luo, Videtto, Mancuso, and
  Walker]{Grugan2012}
P.~D. Grugan, S.~Luo, M.~Videtto, C.~Mancuso, and B.~C. Walker.
\newblock {Classical study of ultrastrong nonperturbative-field interactions
  with a one-electron atom: Validity of the dipole approximation for the
  bound-state interaction}.
\newblock \emph{Phys. Rev. A - At. Mol. Opt. Phys.}, 85\penalty0 (5):\penalty0
  1--8, 2012.
\newblock ISSN 10502947.
\newblock \doi{10.1103/PhysRevA.85.053407}.

\bibitem[Milosevic et~al.(2002{\natexlab{a}})Milosevic, Krainov, and
  Brabec]{Milosevic2002}
N~Milosevic, V~P Krainov, and T~Brabec.
\newblock {Semiclassical Dirac Theory of Tunnel Ionization}.
\newblock \emph{Phys. Rev. Lett.}, 89\penalty0 (19):\penalty0 193001--1 --
  193001--4, 2002{\natexlab{a}}.
\newblock \doi{10.1103/PhysRevLett.89.193001}.

\bibitem[Milosevic et~al.(2002{\natexlab{b}})Milosevic, Krainov, and
  Brabec]{Milosevic2002a}
N.~Milosevic, V.~P. Krainov, and T.~Brabec.
\newblock {Relativistic theory of tunnel ionization}.
\newblock \emph{J. Phys. B At. Mol. Opt. Phys.}, 35\penalty0 (16):\penalty0
  3515--3529, 2002{\natexlab{b}}.
\newblock ISSN 09534075.
\newblock \doi{10.1088/0953-4075/35/16/311}.

\bibitem[Rus et~al.(2017)Rus, Bakule, Kramer, Naylon, Thoma, Fibrich, Green,
  Lagron, Antipenkov, Barton{\'{i}}{\v{c}}ek, Batysta, Ba{\v{s}}e, Boge, Buck,
  Cupal, Drouin, Ďur{\'{a}}k, Himmel, Havl{\'{i}}{\v{c}}ek, Homer, Honsa,
  Hor{\'{a}}{\v{c}}ek, Hr{\'{i}}bek, Hub{\'{a}}{\v{c}}ek, Hubka, Kalinchenko,
  Kasl, Indra, Korous, Ko{\v{s}}elja, Koub{\'{i}}kov{\'{a}}, Laub, Mazanec,
  Meadows, Nov{\'{a}}k, Peceli, Polan, Snopek, {\v{S}}obr, Trojek, Tykalewicz,
  Velpula, Verhagen, Vyhl{\'{i}}dka, Weiss, Haefner, Bayramian, Betts,
  Erlandson, Jarboe, Johnson, Horner, Kim, Koh, Marshall, Mason, Sistrunk,
  Smith, Spinka, Stanley, Stolz, Suratwala, Telford, Ditmire, Gaul, Donovan,
  Frederickson, Friedman, Hammond, Hidinger, Ch{\'{e}}riaux, Jochmann, Kepler,
  Malato, Martinez, Metzger, Schultze, Mason, Ertel, Lintern, Edwards,
  Hernandez-Gomez, and Collier]{Rus2017}
B.~Rus, P.~Bakule, D.~Kramer, J.~Naylon, J.~Thoma, M.~Fibrich, J.~T. Green,
  J.~C. Lagron, R.~Antipenkov, J.~Barton{\'{i}}{\v{c}}ek, F.~Batysta,
  R.~Ba{\v{s}}e, R.~Boge, S.~Buck, J.~Cupal, M.~A. Drouin, M.~Ďur{\'{a}}k,
  B.~Himmel, T.~Havl{\'{i}}{\v{c}}ek, P.~Homer, A.~Honsa,
  M.~Hor{\'{a}}{\v{c}}ek, P.~Hr{\'{i}}bek, J~Hub{\'{a}}{\v{c}}ek, Z.~Hubka,
  G.~Kalinchenko, K.~Kasl, L.~Indra, P.~Korous, M.~Ko{\v{s}}elja,
  L.~Koub{\'{i}}kov{\'{a}}, M.~Laub, T.~Mazanec, A.~Meadows, J.~Nov{\'{a}}k,
  D.~Peceli, J.~Polan, D.~Snopek, V.~{\v{S}}obr, P.~Trojek, B.~Tykalewicz,
  P.~Velpula, E.~Verhagen, {\v{S}}.~Vyhl{\'{i}}dka, J.~Weiss, C.~Haefner,
  A.~Bayramian, S.~Betts, A.~Erlandson, J.~Jarboe, G.~Johnson, J.~Horner,
  D.~Kim, E.~Koh, C.~Marshall, D.~Mason, E.~Sistrunk, D.~Smith, T.~Spinka,
  J.~Stanley, C.~Stolz, T.~Suratwala, S.~Telford, T.~Ditmire, E.~Gaul,
  M.~Donovan, C.~Frederickson, G.~Friedman, D.~Hammond, D.~Hidinger,
  G.~Ch{\'{e}}riaux, A.~Jochmann, M.~Kepler, C.~Malato, M.~Martinez,
  T.~Metzger, M.~Schultze, P.~Mason, K.~Ertel, A.~Lintern, C.~Edwards,
  C.~Hernandez-Gomez, and J.~Collier.
\newblock {ELI-beamlines: progress in development of next generation
  short-pulse laser systems}.
\newblock \emph{Res. Using Extrem. Light Enter. New Front. with Petawatt-Class
  Lasers III}, 10241\penalty0 (May 2015):\penalty0 102410J, 2017.
\newblock \doi{10.1117/12.2269818}.

\bibitem[Salamin et~al.(2008)Salamin, Harman, and Keitel]{Salamin2008}
Yousef~I. Salamin, Zolt{\'{a}}n Harman, and Christoph~H. Keitel.
\newblock {Direct high-power laser acceleration of ions for medical
  applications}.
\newblock \emph{Phys. Rev. Lett.}, 100\penalty0 (15):\penalty0 1--4, 2008.
\newblock ISSN 00319007.
\newblock \doi{10.1103/PhysRevLett.100.155004}.

\bibitem[Ammosov et~al.(1986)Ammosov, Delone, and Krainov]{Ammosov1986}
M.~V. Ammosov, N.~B. Delone, and V.~P. Krainov.
\newblock {Tunnel ionization of complex atoms and of atomic ions in an
  alternating electromagnetic field}.
\newblock \emph{J. Exp. Theor. Phys.}, 64\penalty0 (6):\penalty0 1191 -- 1194,
  1986.
\newblock ISSN 02635747.
\newblock \doi{10.1017/S0263574703005605}.

\bibitem[Perelomov et~al.(1966)Perelomov, Popov, and Terent'ev]{Perelomov1966}
A.~Perelomov, V.~Popov, and M.~Terent'ev.
\newblock {Ionization of Atoms in an Alternating Electric Field}.
\newblock \emph{Sov. J. Exp. Theor. Phys.}, 23\penalty0 (5):\penalty0 924,
  1966.
\newblock ISSN 1063-7761.

\bibitem[Kelly and Harrison(1971)]{Kelly1971}
Raymond~L Kelly and Don E.~Jr Harrison.
\newblock {Ionization potentials, experimental and theoretical, of the elements
  hydrogen to krypton}.
\newblock \emph{At. Data Nucl. Data Tables}, 3:\penalty0 177--193, 1971.
\newblock ISSN 2047-8844.
\newblock \doi{10.1111/2047-8852.12112}.

\bibitem[Kramida et~al.(2018)Kramida, {Yu.~Ralchenko}, Reader, and {and NIST
  ASD Team}]{NIST_ASD}
A.~Kramida, {Yu.~Ralchenko}, J.~Reader, and {and NIST ASD Team}.
\newblock {NIST Atomic Spectra Database (ver. 5.6.1), [Online]. Available:
  {\tt{https://physics.nist.gov/asd}} [2019, October 2]. National Institute of
  Standards and Technology, Gaithersburg, MD.}, 2018.

\bibitem[Salamin(2007)]{Salamin2007}
Y.I. Salamin.
\newblock {Fields of a Gaussian beam beyond the paraxial approximation}.
\newblock \emph{Appl. Phys. B}, 86:\penalty0 319--326, 2007.
\newblock \doi{10.1007/s00340-006-2442-4}.

\bibitem[Augst et~al.(1991)Augst, Meyerhofer, Strickland, and Chin]{Augst2008}
S.~Augst, D.~D. Meyerhofer, D.~Strickland, and S.~L. Chin.
\newblock {Laser ionization of noble gases by Coulomb-barrier suppression}.
\newblock \emph{J. Opt. Soc. Am. B}, 8\penalty0 (4):\penalty0 858--867, 1991.
\newblock ISSN 0740-3224.
\newblock \doi{10.1364/josab.8.000858}.

\bibitem[Bauer and Mulser(1999)]{Bauer1999}
D~Bauer and P~Mulser.
\newblock {Exact field ionization rates in the barrier-suppression regime from
  numerical time-dependent Schrodinger-equation calculations}.
\newblock \emph{Phys. Rev. A - At. Mol. Opt. Phys.}, 59\penalty0 (1):\penalty0
  569--577, 1999.

\bibitem[Ciappina et~al.(2019)Ciappina, Popruzhenko, Bulanov, Ditmire, Korn,
  and Weber]{Ciappina2019}
M.~F. Ciappina, S.~V. Popruzhenko, S.~V. Bulanov, T.~Ditmire, G.~Korn, and
  S.~Weber.
\newblock {Progress toward atomic diagnostics of ultrahigh laser intensities}.
\newblock \emph{Phys. Rev. A}, 99\penalty0 (4):\penalty0 1--13, 2019.
\newblock ISSN 24699934.
\newblock \doi{10.1103/PhysRevA.99.043405}.

\bibitem[Kostyukov and Golovanov(2018)]{Kostyukov2018}
I.~Yu Kostyukov and A.~A. Golovanov.
\newblock {Field ionization in short and extremely intense laser pulses}.
\newblock \emph{Phys. Rev. A}, 98\penalty0 (4), 2018.
\newblock ISSN 24699934.
\newblock \doi{10.1103/PhysRevA.98.043407}.

\bibitem[Bucksbaum et~al.(1987)Bucksbaum, Freeman, Bashkansky, and
  McIlrath]{Bucksbaum2008}
P.~H. Bucksbaum, R.~R. Freeman, M.~Bashkansky, and T.~J. McIlrath.
\newblock {Role of the ponderomotive potential in above-threshold ionization}.
\newblock \emph{J. Opt. Soc. Am. B}, 4\penalty0 (5):\penalty0 760--764, 1987.
\newblock ISSN 0740-3224.
\newblock \doi{10.1364/josab.4.000760}.

\bibitem[Gordon et~al.(2017)Gordon, Palastro, and Hafizi]{Gordon2017}
D.~F. Gordon, J.~P. Palastro, and B.~Hafizi.
\newblock {Superponderomotive regime of tunneling ionization}.
\newblock \emph{Phys. Rev. A - At. Mol. Opt. Phys.}, 95\penalty0 (3):\penalty0
  1--5, 2017.
\newblock ISSN 24699934.
\newblock \doi{10.1103/PhysRevA.95.033403}.

\bibitem[Pi et~al.(2015)Pi, Hu, and Starace]{Pi2015}
Liang~Wen Pi, S.~X. Hu, and Anthony~F. Starace.
\newblock {Favorable target positions for intense laser acceleration of
  electrons in hydrogen-like, highly-charged ions}.
\newblock \emph{Phys. Plasmas}, 22\penalty0 (9), 2015.
\newblock ISSN 10897674.
\newblock \doi{10.1063/1.4930218}.

\bibitem[Prince and Dormand(1981)]{Prince1981}
P~J Prince and J~R Dormand.
\newblock {High order embedded Runge-Kutta formulae}.
\newblock \emph{J. Comput. Appl. Math.}, 7\penalty0 (1):\penalty0 67--75, 1981.

\bibitem[Tamburini et~al.(2010)Tamburini, Pegoraro, {Di Piazza}, Keitel, and
  Macchi]{Tamburini2010}
M.~Tamburini, F.~Pegoraro, A.~{Di Piazza}, C.~H. Keitel, and A.~Macchi.
\newblock {Radiation reaction effects on radiation pressure acceleration}.
\newblock \emph{New J. Phys.}, 12\penalty0 (123005), 2010.
\newblock \doi{10.1088/1367-2630/12/12/123005}.

\bibitem[Tamburini et~al.(2012)Tamburini, Liseykina, Pegoraro, and
  MacChi]{Tamburini2012}
M.~Tamburini, T.~V. Liseykina, F.~Pegoraro, and A.~MacChi.
\newblock {Radiation-pressure-dominant acceleration: Polarization and radiation
  reaction effects and energy increase in three-dimensional simulations}.
\newblock \emph{Phys. Rev. E - Stat. Nonlinear, Soft Matter Phys.}, 85\penalty0
  (1):\penalty0 1--5, 2012.
\newblock ISSN 15393755.
\newblock \doi{10.1103/PhysRevE.85.016407}.

\bibitem[Quesnel and Mora(1998)]{Quesnel1998}
Brice Quesnel and Patrick Mora.
\newblock {Theory and Simulation of the Interaction of Ultraintense Laser Pules
  with Electrons in Vacuum}.
\newblock \emph{Phys. Rev. E - Stat. Nonlinear, Soft Matter Phys.}, 58\penalty0
  (3):\penalty0 1--14, 1998.
\newblock URL
  \url{papers2://publication/uuid/D507B656-2A24-4BAD-A590-644268AA54EE}.

\bibitem[Maltsev and Ditmire(2003)]{Maltsev2003a}
A.~Maltsev and T.~Ditmire.
\newblock {Above Threshold Ionization in Tightly Focused, Strongly Relativistic
  Laser Fields}.
\newblock \emph{Phys. Rev. Lett.}, 90\penalty0 (5):\penalty0 4, 2003.
\newblock ISSN 10797114.
\newblock \doi{10.1103/PhysRevLett.90.053002}.

\bibitem[Popov et~al.(2008)Popov, Bychenkov, Rozmus, and Sydora]{Popov2008}
K.~I. Popov, V.~Yu Bychenkov, W.~Rozmus, and R.~D. Sydora.
\newblock {Electron vacuum acceleration by a tightly focused laser pulse}.
\newblock \emph{Phys. Plasmas}, 15\penalty0 (1), 2008.
\newblock ISSN 1070664X.
\newblock \doi{10.1063/1.2830651}.

\bibitem[Marceau et~al.(2012)Marceau, April, and Pich{\'{e}}]{Marceau2012}
Vincent Marceau, Alexandre April, and Michel Pich{\'{e}}.
\newblock {Electron acceleration driven by ultrashort and nonparaxial radially
  polarized laser pulses}.
\newblock \emph{Opt. Lett.}, 37\penalty0 (13):\penalty0 2442, 2012.
\newblock ISSN 0146-9592.
\newblock \doi{10.1364/ol.37.002442}.

\bibitem[Augst et~al.(1989)Augst, Strickland, Meyerhofer, Chin, and
  Eberly]{Augst1989}
S.~Augst, D.~Strickland, D.~D. Meyerhofer, S.~L. Chin, and J.~H. Eberly.
\newblock {Tunneling ionization of noble gases in a high-intensity laser
  field}.
\newblock \emph{Phys. Rev. Lett.}, 63\penalty0 (20):\penalty0 2212--2215, 1989.
\newblock ISSN 00319007.
\newblock \doi{10.1103/PhysRevLett.63.2212}.

\bibitem[Akahane et~al.(2006)Akahane, Ma, Fukuda, Aoyoma, Kiriyama, Sheldakova,
  Kudryashov, and Yamakawa]{Akahane2006}
Yutaka Akahane, Jinglong Ma, Yuji Fukuda, Makoto Aoyoma, Hiromitsu Kiriyama,
  Julia~V. Sheldakova, Alexis~V. Kudryashov, and Koichi Yamakawa.
\newblock {Characterization of wave-front corrected 100 TW, 10 Hz laser pulses
  with peak intensities greater than $10^{20}$ W/cm$^{2}$}.
\newblock \emph{Rev. Sci. Instrum.}, 77\penalty0 (2), 2006.
\newblock ISSN 00346748.
\newblock \doi{10.1063/1.2166669}.
\newblock URL \url{https://doi.org/10.1063/1.2166669}.

\bibitem[Link et~al.(2006)Link, Chowdhury, Morrison, Ovchinnikov, Offermann,
  {Van Woerkom}, Freeman, Pasley, Shipton, Beg, Rambo, Schwarz, Geissel, Edens,
  and Porter]{Link2006}
Anthony Link, Enam~A. Chowdhury, John~T. Morrison, Vladimir~M. Ovchinnikov,
  Dustin Offermann, Linn {Van Woerkom}, Richard~R. Freeman, John Pasley, Erik
  Shipton, Farhat Beg, Patrick Rambo, Jens Schwarz, Matthias Geissel, Aaron
  Edens, and John~L. Porter.
\newblock {Development of an in situ peak intensity measurement method for
  ultraintense single shot laser-plasma experiments at the Sandia Z petawatt
  facility}.
\newblock \emph{Rev. Sci. Instrum.}, 77\penalty0 (10), 2006.
\newblock ISSN 00346748.
\newblock \doi{10.1063/1.2336469}.
\newblock URL \url{https://doi.org/10.1063/1.2336469}.

\bibitem[Dichiara et~al.(2008)Dichiara, Ghebregziabher, Sauer, Waesche,
  Palaniyappan, Wen, and Walker]{Dichiara2008}
A.~D. Dichiara, I.~Ghebregziabher, R.~Sauer, J.~Waesche, S.~Palaniyappan, B.~L.
  Wen, and B.~C. Walker.
\newblock {Relativistic MeV photoelectrons from the single atom response of
  argon to a $10^{19}$ W/cm$^{2}$ laser field}.
\newblock \emph{Phys. Rev. Lett.}, 101\penalty0 (17):\penalty0 1--4, 2008.
\newblock ISSN 00319007.
\newblock \doi{10.1103/PhysRevLett.101.173002}.

\bibitem[Ekanayake et~al.(2013)Ekanayake, Luo, Grugan, Crosby, Camilo, Mccowan,
  Scalzi, Tramontozzi, Howard, Wells, Mancuso, Stanev, Decamp, and
  Walker]{Ekanayake2013}
N~Ekanayake, S~Luo, P~D Grugan, W~B Crosby, A~D Camilo, C~V Mccowan, R~Scalzi,
  A~Tramontozzi, L~E Howard, S~J Wells, C~Mancuso, T~Stanev, M~F Decamp, and
  B~C Walker.
\newblock {Electron Shell Ionization of Atoms with Classical , Relativistic
  Scattering}.
\newblock \emph{Phys. Rev. Lett.}, 110\penalty0 (13):\penalty0 1--5, 2013.
\newblock \doi{10.1103/PhysRevLett.110.203003}.

\end{thebibliography}

\end{document}